\documentclass[journal=jacsat,manuscript=article]{achemso}

\usepackage[version=3]{mhchem} 

\author{Haifeng Ma}
\author{Thomas Brugger}
\author{Simon Berner}
\affiliation[Physik-Institut, Universit\"at Z\"urich]
{Physik-Institut}
\author{Yun Ding}
\author{Marcella Iannuzzi}
\author{J\"urg  Hutter}
\affiliation[Physikalisch-Chemisches Institut, Universit\"at Z\"urich]
{Physikalisch-Chemisches Institut, Universit\"at Z\"urich, Winterthurerstrasse 190, CH-8057 Z\"urich, Switzerland}
\author{J\"urg Osterwalder}
\author{Thomas Greber}
\email{greber@physik.uzh.ch}
\affiliation[Physik-Institut, Universit\"at Z\"urich]
{Physik-Institut}

\title[]{Boron Nitride Nanomesh: A template for Nano-ice}

\begin{document}

\begin{abstract}
{Using variable temperature scanning tunneling microscopy and $dI/dz$ barrier height spectroscopy, the structure of water on $h$-BN/Rh(111) nanomesh has been investigated.  Below its desorption temperature, two distinct phases of water self-assemble within the 3.2 nm unit cell of the nanomesh. In the 2 nm holes, an ordered phase of nano-ice crystals with about 40 molecules is found. The ice crystals arrange in a bilayer honeycomb lattice, where the hydrogen atoms of the lower layer point to the substrate.
The phase on the 1 nm wires, is a low density gas phase, which is characterized by contrast modulations and streaky noise in the STM images. 
Tunneling barrier measurements infer the proton positions in the nano-ice clusters. }
\end{abstract}
\date{\today}
\section{Introduction}
{
Templates are essential in nanoscience and technology, since they allow the self-assembly of seemingly artificial structures with new functionalities.
If a surface shall act as a natural template that imposes a lateral ordering on a length scale larger than a 1$\times$1 unit cell, it has to rely on reconstruction and the formation of superstructures. 
For the case of the $h$-BN/Rh(111) nanomesh \cite{cor04} the superstructure bases on 12$\times$12 Rh unit cells on top of which 13$\times$13 BN unit cells coincide. \cite{las07,ber07}
It is a corrugated single layer of sp$^2$ hybridized hexagonal boron nitride with a lattice constant of 3.2 nm. 
If the nitrogen atom sits close to an on top position of a substrate Rh atom, $h$-BN is tightly bound and forms the so called "holes" or "pores" with a diameter of about 2 nm. The loosely bound regions form a connected network and are called wires.
This structure imposes a template function which can be used for trapping single molecules at room temperature.\cite{ber07}
The trapping mechanism was found to be related to the presence of large lateral electric fields in the surface, where the rims of the holes carry lateral dipole rings, which produce a lower electrostatic potential in the holes. \cite{dil08} 
Also, it was shown that the $h$-BN nanomesh is very robust and survives transport in air,\cite{bun07} immersion into liquids \cite{ber07}, and even electrolytic cycles. \cite{wid07} 
The impedance analysis in the electrochemical set up suggested that the $h$-BN nanomesh imposes ordering in the electrolyte within the Helmholtz layer. \cite{wid07}
This led us to the investigation of water on the nanomesh. 
It turns out that the hydrogen bonds dominate above the bonds due to the dipole rings. However, the rims of the holes prevent the ice clusters from further growth. It also becomes apparent that the wires act like a feeding/draining network that connects the ice clusters. 

Water on surfaces is an intriguing subject with many experimental \cite{hen02,thi87,ver06,son09,ron02,mor07} and theoretical \cite{feibelman02,mic04,saa05,pan08,tos08} challenges. In this letter the $h$-BN/Rh(111) nanomesh is used as a template for the formation of ice clusters. It is found that at temperatures below 50 K water self-assembles in the nanomesh holes into two dimensional ice clusters with about 40 molecules, and remains as a low density phase on the wires. The structure of the clusters is elucidated, and in particular it is shown that the positions of the hydrogen atoms within the clusters may be inferred from tunneling barrier measurements with sub-nanometer resolution.
    

\section{Experimental}
{The experiments were performed in an ultrahigh-vacuum (UHV) chamber with a background pressure below 4$\times10^{-10}$ Torr using variable-temperature scanning tunneling microscopy (Omicron VT-STM). The sample preparation included several cycles of Ar$^{+}$-bombardment, subsequent exposure to a few L (1 Langmuir=10$^{-6}$ Torr$\cdot$s) of O$_2$ and annealing of the Rh(111) sample. Then it was exposed to 40 L of borazine (HBNH)$_3$ gas, while keeping the surface at 1070 K. This procedure yields a well-ordered large-scale single layer of hexagonal boron nitride on Rh(111) surface. Milli-Q water was used and purified by several freeze-and-pump cycles. Water was introduced into the UHV chamber via a leak valve in the pressure range of 10$^{-10}$-10$^{-8}$ Torr through a nozzle pointing towards the sample. All pressures correspond to the uncorrected reading of the ion gauge of the UHV chamber. In order to minimize tip induced H$_2$O motion on the surface, scanning parameters were set to the order of 50 pA to 400 pA for tunneling currents and ${-}$50 mV to ${-}$400 mV for tunneling voltages. Tunneling barrier spectroscopy was performed with lock-in technique by superimposing a modulation voltage of 8 mV at 2.0 k$Hz$ to the
$z$-piezo. This modulates the tip-sample distance with an amplitude of $\sim$0.07 nm. Except for those mentioned explicitly, STM images shown in this paper are obtained at low temperature 34 K.
\section{Results and discussion}
\ref{F1} compares the $h$-BN/Rh(111) nanomesh with and without adsorbed water. The images with atomic resolution show that water forms a two-phase system, where the two phases coexist in the supercell with 3.2 nm lattice constant. \ref{F1}a is a large scale STM current image of the nanostructure (42$\times$42 nm$^{2}$) across several terraces of the Rh(111) substrate, obtained after dosing about 2 L water on the surface at 52 K. 
All holes are filled with ice clusters and no ordered water structures appear on the wires. This indicates that in the holes the interaction between water and the surface is strong enough to stabilize molecules. Compared to the empty mesh (\ref{F1}b), where the holes and the wires are visible as two structural elements of the nanomesh, the holes are filled with small lobes and streaks appear along the scan direction after dosing 1 L of water (\ref{F1}c). The topographic cross sections across a hole in \ref{F1}d demonstrate that those water molecules map as protrusions with a height of about 15 pm, which is in agreement with the known mapping properties of water on surfaces.\cite{shi08} The lateral distance between the protrusions is about 0.46 nm in the holes with a remaining depth of about 40 pm. As we show in the following, it is the signature of a bilayer of water --- a lateral honeycomb structure with two sublattices A and B, where molecules in sublattice A form the contrast in the image. 
Furthermore, it is observed that no such ordered ice clusters appear on the wires, though they map rougher and noisy streaks reminiscent to mobile species are noticed.

\ref{F1}e shows that the height differences of the cuts around the wires are irregular, where the height variations are, compared to the empty nanomesh, larger for the water covered case. We speculate that, water molecules are moving fast along wires to form one low density gas phase which is e.g. similar to the 2D solid-gas structure obtained for SubPc molecules on Ag(111) surface.\cite{ber01} These results indicate that the water self-assembly process in the boron nitride nanomesh, separates water into two distinct phases, one on the wires as fuzzy protrusions and one in the holes of the template as ordered nano-ice crystal. In the following we focus on the structure of the ordered ice clusters in the holes.

The lattice constants and the azimuthal orientation can be seen in  \ref{F2}.  
In  \ref{F2}a  and c, STM pictures of two different preparations are shown. 
The Fourier transforms (\ref{F2}b  and d) reveal the two characteristic lattice constants of the nanomesh (3.2 nm) and the ice clusters ($\sim$0.46 nm). The lattice constant of ice is 7.0$\pm$0.2 times smaller than that of the nanomesh and corresponds within the error bar to that of the basal plane of hexagonal ice (0.45 nm). For the two shown preparations we find, different orientations rotated by 11$\pm$3$\,^{\circ}$  and -24$\pm$3$\,^{\circ}$ with respect to the direction given by two neighboring holes (\ref{F2}b and d).
The observation of rigid ice clusters that are not oriented the same way with respect to the substrate indicate that the lock in energy of the nano-ice clusters is not as large as the bonding within the ice layer. 
The histogram in \ref{F2}e lists the number of molecular protrusions in the 32 holes of  \ref{F2}a where in average 20$\pm$5 protrusions per hole were found. This confirms the above statement that water in sublattice A forms the contrast, since a hole with 2.2 nm diameter is expected to host twice as many i.e. about 40 water molecules per bilayer. 

\ref{F3} substantiates the two sublattice (bilayer) picture with a high resolution zoom into an ice cluster. The two sublattices A and B were resolved. We assign the highest protrusions A to positions of the topmost oxygen atoms, while the protrusions with intermediate height indicate oxygen atoms of sublattice B. From the cross section in \ref{F3}b it can be seen that sublattice B appears at the given tunneling conditions about 15 pm below sublattice A, where the direction ${\overline{\rm AB}}$ between nearest neighbor oxygen molecules encloses an angle of 93$^\circ$ with the surface normal.  This confirms that at 34 K ice crystallizes in a bilayer structure in the holes of the $h$-BN/Rh(111) nanomesh. While it is intuitive to assign the two sublattices to two differently bound water molecules, the topographic images make no direct statement on the positions of the hydrogen atoms, nor the orientation of the lonepairs on the oxygen atoms. 
In the following we show that the water dipoles and thus the positions of the hydrogen atoms affect the tunneling barrier. For this purpose models for the bilayer that obey the ice rules are considered. Essentially the ice rules say that every hydrogen atom forms a hydrogen bond with a neighboring oxygen atom. Every water molecule is modeled by a tetrahedron, with an oxygen atom in the center, two hydrogen atoms or protons in two of its vertices, and two oxygen lone pairs in the two remaining vertices of the tetrahedron.\cite{hen02}. For such a hexagonal bilayer the angle between ${\overline{\rm{AB}}}$ and the surface normal is expected to be the tetrahedron angle of 109$^\circ$. The deviation to the observed 93$^\circ$ is merely taken as an indication that the apparent heights in STM pictures must not correspond to the true topography, since it is also influenced by the local density of states and the tunneling barrier and that the tetrahedra might be distorted perpendicular to the surface. More importantly, the tetrahedra model does not make a statement on which vertices the protons sit.
The basal hexagram plane is fully hydrogen bonded, i.e. half of the vertices are occupied with hydrogen atoms, and their nearest neighbor vertices are empty. It is worth to recall that the ice rules allow many different arrangements of the protons, i.e. allow for a certain proton disorder.
The up (down) vertices either host a hydrogen (lone pair) or vice versa.
The issue on whether hydrogen atoms in sublattice A (B) point to the substrates or not, can be addressed by  barrier height $dI/dz$ microscopy \cite{vri08,qiy07}, since the positions of the ions leave fingerprints in the electrostatic landscape, as it was e.g. shown with image potential state and $dI/dV$ spectroscopy for NaCl layers \cite{piv05,ols05}.
Electrostatic potential variations locally affect the tunneling barrier height. The tunneling barrier height is determined in measuring the $z$-dependence of the tunneling current. The tunneling current $I$ corresponds exponentially on the distance $z$  between the tip and the sample: $I \propto \exp(-z/z_o)$, where the tunneling length $z_o$ depends on the tunneling barrier. 
For small tunneling voltages $z_o$ is roughly proportional to  $1/\sqrt{\Phi}_l$, where $\Phi_l$ is the local work function in the tunneling junction. In the barrier height microscopy mode the $z$-position of the tip is oscillated with a given frequency, while the tunneling current variations at this frequency are read out, and related via $(d\ln I/dz)^2\propto \Phi_l$ to the local work function \cite{vri08}.
\ref{F4}a shows a corresponding $dI/dz$ map of nano-ice/$h$-BN/Rh(111). 
Clearly, at the positions of the oxygen atoms of sublattice A and on the wires, a larger tunneling barrier is found, where there is some scatter in the barrier height above the individual up vertices. 
This is consistent with the assignment that the up vertices of sublattice A host oxygen lone pairs and correspondingly that the out of plane hydrogen atoms sit on the water molecules of sublattice B and point to the $h$-BN substrate.
Density functional theory calculations of single water molecules on a single sheet of $h$-BN confirm this picture. They favor single water molecules to bind with the hydrogen to the negatively charged nitrogen atoms. 

For comparison we also show the $dI/dz$ map of a bare nanomesh hole and the confining wires (\ref{F4}b). 
This result with atomic resolution on the $h$-BN is in perfect agreement with the potential energy landscape of the clean nanomesh \cite{dil08}. 
Both pictures in \ref{F4}a and b are recorded with the same tunneling parameters. The magnitude of the effect may be estimated from the measured potential energy difference 0.38 nm above the nitrogen atoms of 310 meV between holes and wires in the empty nanomesh \cite {dil08}.

In order to substantiate the $dI/dz$ maps we built bilayer models of ice clusters obeying the above mentioned ice rules. 
A model of such an ice cluster is shown in \ref{F4}c. It shows 42 water molecules arranged in a honeycomb fashion, where in sublattice B the hydrogen atoms point to the substrate. There are several possibilities to arrange the hydrogen atoms. Here they are arranged in such a way that the cluster has no lateral dipolemoment. From density functional theory we take the Mullikan charges on the oxygen atoms of 0.4 electrons and ${-}$0.2 electrons on the hydrogen atoms (protons). With this we calculate the electrostatic potential above the cluster in \ref{F4}c. \ref{F4}d shows this potential 0.2 nm above the plane between sublattice A and B. Clearly, the potential that is reminiscent to a local work function is largest on top of the oxygen atoms in sublattice A, i.e. on top of the lone pairs that are not saturated with hydrogen atoms. For the chosen parameters we find a corrugation of the electrostatic potential energy landscape of  up to 0.8 eV within the cluster. It has to be noted that this energy corrugation is distance dependent, and reproduces the order of magnitude of the effect. The calculations of the tunneling barrier also indicate that the $dI/dz$ measurements feature the electrostatic potential closer to the surface than adsorbed Xe (0.38 nm) \cite{dil08}, since at this distance the molecular resolution is lost.
All other configurations we modeled, like e.g. a structure with hydrogen atoms pointing to the vacuum, do not produce this kind of agreement. It is interesting to point out that the variations of the tunneling barrier heights on the different water molecules in sublattice A indicate that the method is sensitive to the local arrangement of the dipoles of the water molecules and that this  bears the potential to address the positions of the protons.
Also the findings indicate that different proton configurations near the edge of the nano-ice clusters cause electric fields that are important for the understanding of the growth limitation of the clusters and their reaction against other host molecules. 

In conclusion we report on the self-assembly of nano-ice clusters with about 40 molecules and a dilute "gas phase" on the $h$-BN/Rh(111) nanomesh template. The nanomesh holes host the clusters and limit their size. With the help of barrier height $dI/dz$ spectroscopy the structure of the nano-ice could be refined to that of a hydrogen bonded honeycomb bilayer, where the out of plane hydrogen atoms bond toward the surface. Also the positions of the protons in the layer may be inferred from the orientations of the water dipole fields. This new structure of ice gives unprecedented insight into the self assembly process of water and indicates it to be a good candidate to study proton disorder in two dimensional ice.
\begin{acknowledgement}

The authors thank Ari Seitsonen and Heinz Blatter for fruitful discussions and Martin Kl\"ockner for skillful technical assistance. 
The project is supported by Sinergia program of the Swiss National Science Foundation.


\end{acknowledgement}




\bibliography{references}

\begin{figure}
\includegraphics[width=\textwidth]{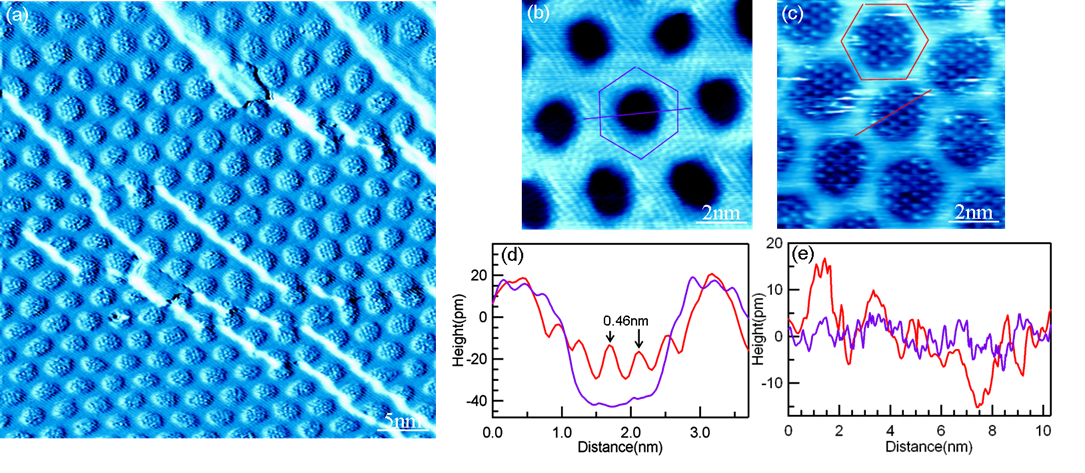}
\caption{\label{F1}Nano-ice on the h-BN/Rh(111) surface. (a) Large-scale STM current image of water molecules adsorbed on the surface at T=52 K. All holes are occupied by distinct ice clusters (42 nm$\times$42 nm; V$_s$= -0.4 V; I$_t$= 400 pA). (b) Atomically-resolved STM topography images of the bare surface and after dosing water (c), both areas are 9 nm$\times$9 nm; and scanning parameters are V$_s$= -0.05 V; I$_t$= 100 pA. Height profiles across (d) and around (e) the holes in (b) and (c) (blue lines and red lines, respectively). The height level 0 in (d) corresponds to the average of (c) and that in (e) to its average.}
\end{figure}
\begin{figure}
\includegraphics[width=0.5\textwidth]{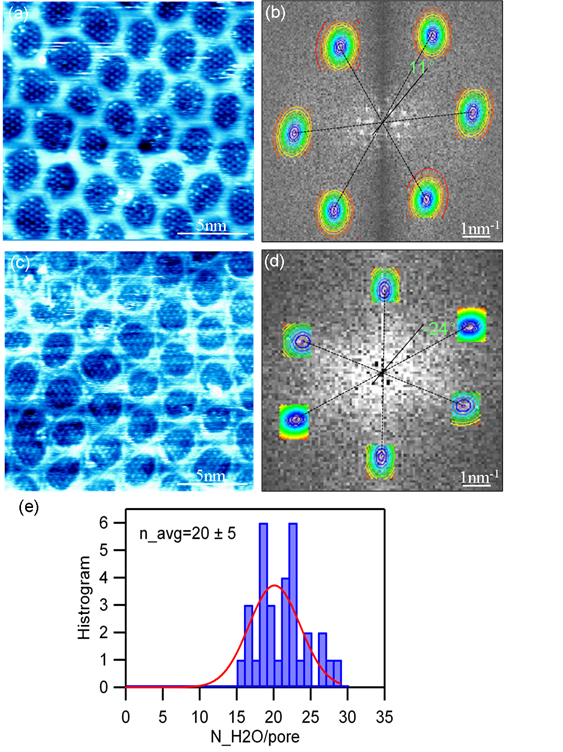}
\caption{\label{F2}Two different preparations of nano-ice clusters on the surface. (a) and (c) STM topography images of nano-ice on h-BN/Rh(111) nanomesh (18 nm$\times$18 nm). (b) and (d) The corresponding Fourier transforms of (a) and (c), respectively. The contours of a Gaussian anchor the exact position for the periodicity of water molecules in reciprocal space. The inset angles corresponding to the orientations between the direction of nano-ice and its neighboring holes. (e) The histogram of the numbers of top layer H$_2$O molecules per hole for the area in (a).}
\end{figure}

\begin{figure}
\includegraphics[width=0.5\textwidth]{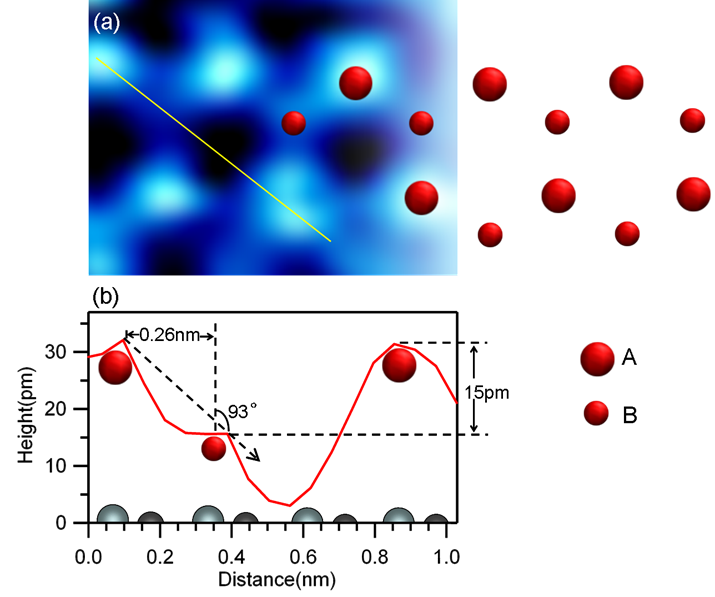}
\caption{\label{F3}Atomic-scale structure of the nano-ice cluster revealing the honeycomb bilayer with two sublattices A and B. (a) High-resolution STM image with the positions of the oxygen atoms (red balls A and B) are embedded (0.8 nm$\times$1.2 nm; V$_s$= +0.3 V; I$_t$= 100 pA). (b) Height profile across the water molecules in (a) as indicated by the yellow line. The two different water molecules are schematically indicated. The angle between ${\overline{\rm{AB}}}$ and the surface normal is 93$^\circ$.  }
\end{figure}

\begin{figure}
\includegraphics[width=0.7\textwidth]{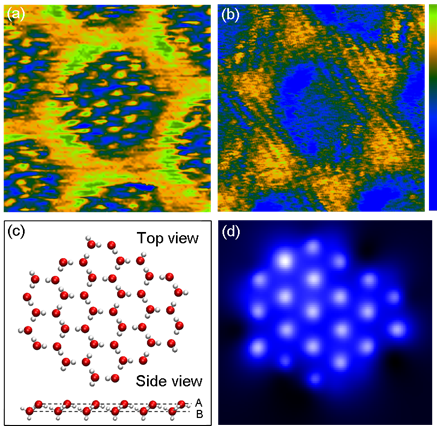}
\caption{\label{F4}(a) The $dI/dz$ map of nano-ice cluster in the hole after dosing 1L H$_2$O molecules on the surface. By measuring the response of the tunneling current to the variation of the tunneling gap distance, $dI/dz$ image can be obtained simultaneously with a topographic STM image. For the $dI/dz$ map, the bright colors correspond to large local tunneling barriers. (b) The $dI/dz$ map was measured on the bare surface (Both images are 6 nm$\times$6 nm; scanning parameters are V$_s$= -0.05 V; I$_t$= 100 pA). (c) Side and top views of the structure model for nano-ice cluster in a single hole. (d) The Corresponding surface electrostatic potential 0.2 nm above the plane between sublattice A and B in (c) (3 nm$\times$3 nm). 
}
\end{figure}

\end{document}